\documentclass[draft]{appolb}
\preprint{hep-ph/0304122}

\usepackage[final]{graphicx}
\usepackage{bm}
\usepackage{citesort}

\newcommand\T[1]{%
  \ensuremath{\bm{#1}_T}
}

\newcommand\query{%
   \fbox{\small ??}%
}

\DeclareRobustCommand{\MSbar}{\ensuremath{ \overline{\rm MS} }}

\begin{document}
\title{What exactly is a parton density?%
}
\author{John C. Collins\thanks{\texttt{collins@phys.psu.edu}}
\address{Physics Department,
        Penn State University,
        104 Davey Laboratory,\\
        University Park PA 16802
        U.S.A.}
}
\maketitle
\begin{abstract}
  I give an account of the definitions of parton densities, both the
  conventional ones, integrated over parton transverse momentum, and
  unintegrated transverse-momentum-dependent densities.  The aim is to
  get a precise and correct definition of a parton density as the
  target expectation value of a suitable quantum mechanical operator,
  so that a clear connection to non-perturbative QCD is provided.
  Starting from the intuitive ideas in the parton model that predate
  QCD, we will see how the simplest operator definitions suffer from
  divergences. Corrections to the definition are needed to eliminate
  the divergences. An improved definition of unintegrated parton
  densities is proposed. 
\end{abstract}

\section{Introduction}
\label{sec:intro}

Central to many of the phenomenological applications of QCD are parton
densities (or parton distribution functions --- pdf's).  The reasons
are quite easy to understand, since the primary tool for making
scattering calculations in quantum field theories is weak coupling
perturbation theory.  Because QCD is a theory of the \emph{strong}
interaction, simple fixed-order perturbation theory is useless for
almost all physical cross sections and amplitudes.  But when a
suitable large momentum scale $Q$ is present, we appeal to
factorization theorems to separate the momentum and distance scales in
a reaction.  The asymptotic freedom of QCD then allows us to use
low-order perturbation theory in powers of $\alpha_s(Q)$ to estimate the
short-distance parts of cross sections.

Pdf's (and related quantities, like fragmentation functions, parton
distribution amplitudes, and generalized parton densities) contain the
non-perturbative parts of the physics.  Because they are universal,
the same pdf's appear in all reactions.  They can be measured in a
limited set of reactions and then perturbative calculations of hard
scattering and pdf evolution enable us to predict, \emph{from first
  principles}, cross sections for many other processes.  The successes
of this formalism are well-known.

A simple example is the factorization theorem for a deep-inelastic
structure function:
\begin{equation}
\label{eq:DIS.factn}
  F_1(x,Q^2) = \sum_i \int_x^1 \frac{d\xi}{\xi}
               C_{1i}(x/\xi, Q^2/\mu^2, \alpha_s(\mu))
               \, f_i(\xi, \mu),
\end{equation}
valid up to power-law corrections at high $Q$.  The standard lowest
order calculation gives $C_{1i}= \frac12 e_i^2\delta(x/\xi-1)$.

Although QCD and factorization appear to form a mature field, the
concepts of parton densities and factorization as presented in much of
the literature are quite problematic.  In fact, as we will see in
Sec.\ \ref{sec:problems}, many of the definitions of pdf's in the
literature, if taken {\em literally}, are wrong.  This should be, and
often is, confusing to students of the subject, even though solutions
to the problems are often well known to experts. Let us just take this
as a symptom of the difficulty of our subject, that of making first
principles predictions from a strongly interacting, relativistic,
quantum many-body theory.

To improve this situation, to make the concepts precise, is
particularly important given the central role that calculations based
on factorization theorems play in extracting physics consequences from
data in high-energy experiments.  Furthermore, as searches for new
physics get more sophisticated, elaborations of the QCD factorization
theorems are needed.  For example, Monte-Carlo event generators are
commonly used to estimate {\em exclusive components} of inclusive
cross sections.  They are generally considered to be theoretically
based on the factorization theorem.  But because a detailed and exact
treatment of parton kinematics is needed, some kind of unintegrated
pdf is needed if the conceptual foundation is to be sound.

In addition, definitions of parton densities form an important link
between treatments of non-perturbative bound states in QCD and their
application, via factorization theorems, to measurable scattering
cross sections.  For this link to work, the definitions must be
correct.

For issues that are not given specific references, the reader can
refer to standard references, for example the book by Ellis, Stirling
and Webber \cite{ESW}, the review of factorization theorems by
Collins, Soper and Sterman \cite{CSS.review}, and the CTEQ handbook
\cite{CTEQ.handbook}.

\section{Development of definitions of pdf's}
\label{sec:problems}

\subsection{Parton model}
\label{sec:parton.model}

\begin{figure}
  \begin{center}
    \includegraphics[scale=0.35]{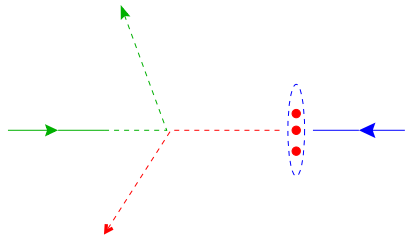}
    \caption{Deep-inelastic scattering.}
    \label{fig:DIS}
  \end{center}
\end{figure}

The basic ideas of pdf's and the parton model are due to Feynman
\cite{Feynman} and predate QCD.  Consider deep-inelastic scattering
(DIS) in the $e$--$p$ center-of-mass frame --- Fig.\ \ref{fig:DIS}.
An electron arrives from the left and undergoes a wide-angle
scattering.  A highly time-dilated and Lorentz contracted proton
arrives from the right; it is symbolized by the squashed blob with 3
dots inside (for the valence quarks).  For numerical illustration,
suppose that $Q^2 = 10^4\,\textrm{GeV}^2$ and $x=0.5$ at the HERA
energy, $\sqrt{s}\simeq 300\,\textrm{GeV}$.  The hard interaction occurs
with one constituent over a scale $1/Q$, about $0.01\,\textrm{fm}$. In
contrast the transverse size of the proton is about $1\,\textrm{fm}$.

The parton model starts from the reasonable supposition that the
interactions binding quarks occur on a time scale $1\,\textrm{fm}/c$
in the rest frame of the proton, and that these get time dilated in
the center-of-mass frame, to about $100\,\textrm{fm}/c$ under the
above conditions.  This suggests that during the interaction of the
electron with the hadronic system, it is a useful {\em approximation}
to assume that the electron interacts with a single fast-moving
constituent, or parton, of the proton, and to neglect the strong
interactions of the parton with the rest of the proton.  That is, the
incoming parton is approximated as a free particle for the purposes of
calculating the interaction with the electron.

Of course, we know that this approximation and the arguments leading
to it are not exactly correct in QCD and other quantum field theories.
Even so, the argument contains a core of truth, and it leads to the
following formula for the structure functions
\begin{equation}
  \label{eq:DIS.parton}
  2xF_1(x,Q^2) = F_2(x,Q^2) =
      \sum_i e_i^2 xf_i(x) + O(\alpha_s(Q)),
\end{equation}
where I have indicated the order of magnitude of the known QCD
corrections (different for $2xF_1$ and $F_2$, of course).

At this level, the pdf $f_i(x)$ is informally defined as the single
particle density of a parton of fractional momentum $x$ and flavor $i$
in a fast moving hadron.  The parton-model formula agrees with the
correct factorization result in QCD when the hard-scattering
coefficient is given its lowest-order approximation and the
DGLAP-evolved pdf's are evaluated at a scale of order $Q$.

\subsection{Light-front quantization}

\begin{figure}
\small
\centering
  \begin{minipage}[t]{0.45\textwidth}
    \centering
    \includegraphics[scale=0.25]{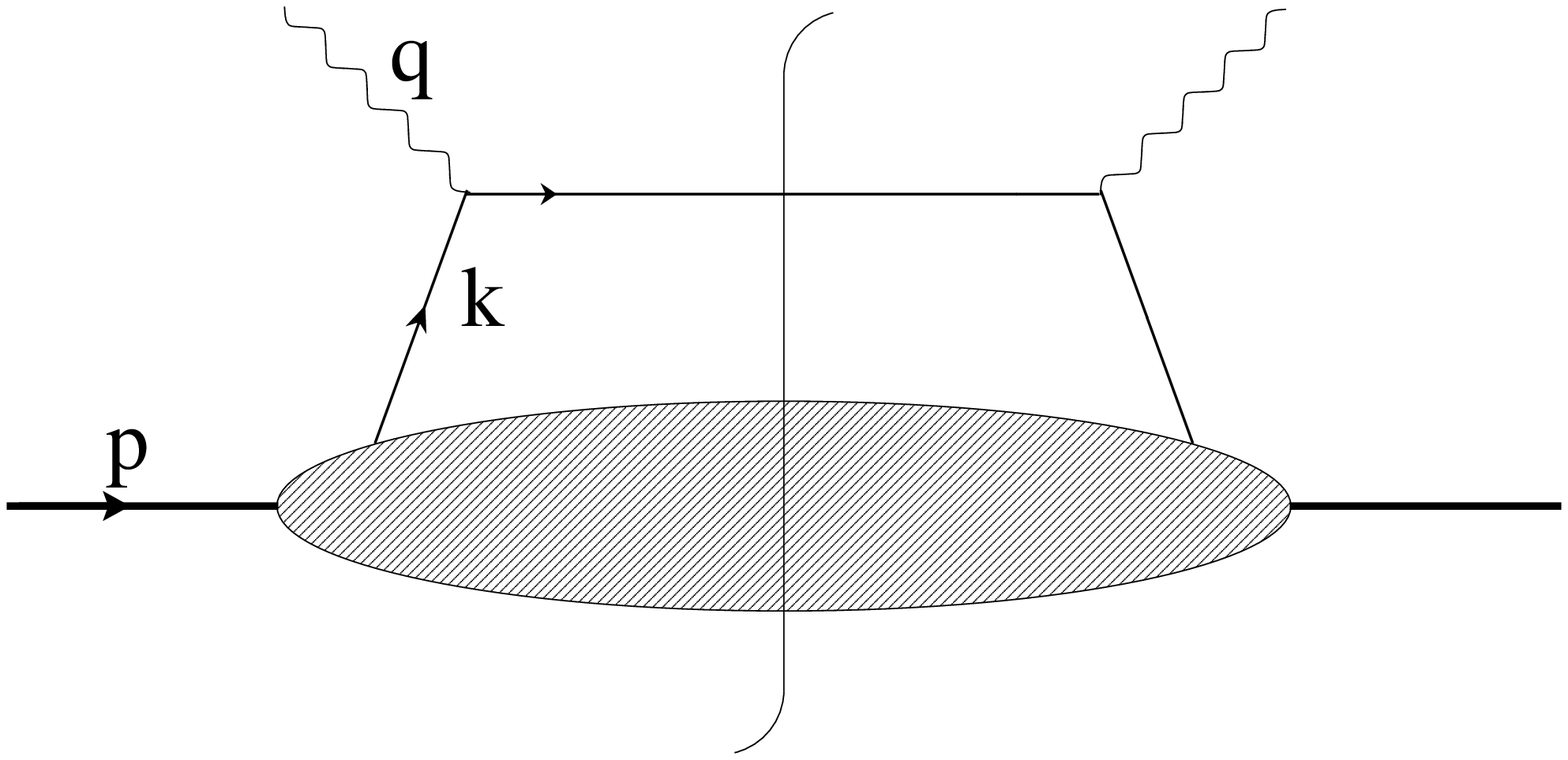}
    \\
    (a)
  \end{minipage}
  \begin{minipage}[t]{0.45\textwidth}
    \centering
    \includegraphics[scale=0.25]{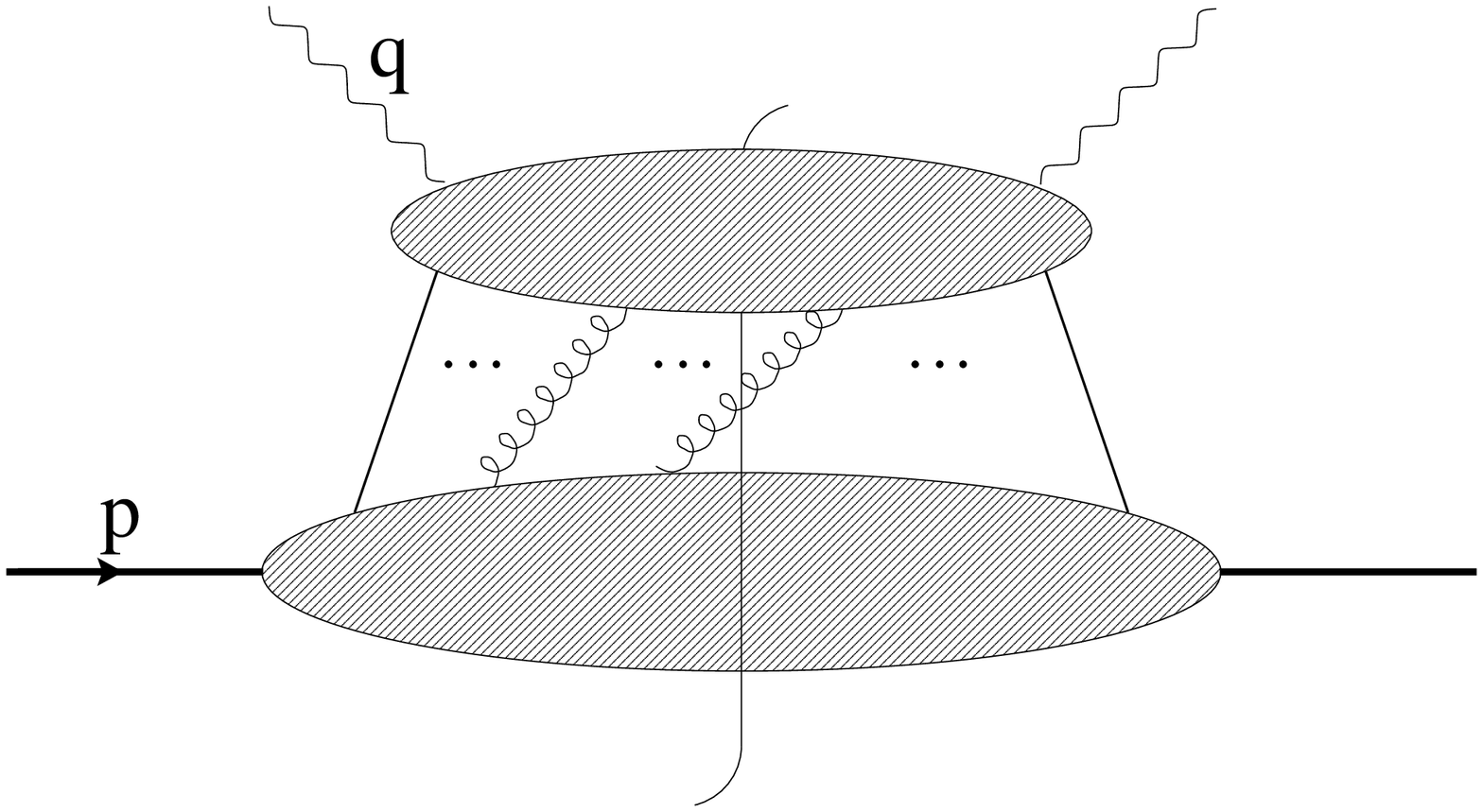}
    \\
    (b)
  \end{minipage}
    \caption{(a) Handbag diagram for DIS.
             (b) Structure of general leading region for DIS.  The
    upper blob has lines with large transverse momentum, and the lower
    blob has lines with low transverse momentum.
    }
    \label{fig:handbag}
\end{figure}

Further progress was made by Bouchiat, Fayet and Meyer \cite{BFM} and
Soper \cite{Soper1,Soper2}.  They observed that the parton model can
be implemented in field theory if one assumes\footnote{These
  assumptions are not exactly correct in QCD, of course.  But the line
  of argument they lead to is useful to explain the definitions of
  pdf's.  } that the dominant contributions have the form of the
``handbag diagram'' of Fig.\ \ref{fig:handbag}(a), and that the
intermediate quark has limited transverse momentum and virtuality.  A
definition of quark pdf's results that can be readily interpreted when
light-front quantization is used to define annihilation and creation
operators $a_i(k^+,\T{k},\lambda)$ and $a_i^{\dag}$ for partons.  Here
$\lambda$ labels the helicity of a parton, and $i$ its flavor.

In this framework, the number density of quarks\footnote{A similar
  definition can be given for the gluon distribution.}, as a function
of fractional longitudinal momentum and transverse momentum is
\cite{Soper2}
\begin{equation}
  \label{eq:pdf.def1}
  P_i(x,\T{k}) = \frac{1}{(2\pi)^3 2x\langle p|p\rangle}
        \langle p| \, {a_i}^{\dag} {a_i}(xp^+,\T{k},\lambda) \, |p\rangle . 
~~\query
\end{equation}
The target of momentum $p^\mu$ is assumed to be moving in the $z$
direction, and the normalization factor follows from the
normalizations of the operators.  The somewhat symbolic division by
$\langle p|p\rangle$ is to be implemented by replacing the state
$|p\rangle$ by a wave packet state, everywhere in Eq.\ 
(\ref{eq:pdf.def1}), and by then taking the limit of a momentum
eigenstate.  In QCD the above definition is not correct as written, as
we will see; this is indicated by the \query{} symbol.  However, since
the parton model and the intuitive space-time picture motivating it
are approximately correct, a correct definition will be structurally
similar.

The pdf can readily be expressed in terms of a quark correlation function:
\begin{equation}
  \label{eq:pdf.def1a}
  P_i(x,\T{k}) =
        \int \frac{dy^-\, d^2\T{y}}{16\pi^3}
           \e^{-ixp^+y^- + i \T{k}\cdot\T{y} }
        \langle p| \, 
           \bar\psi_i(0,y^-, \T{y}) \gamma^+ \psi_i(0)
         \, |p\rangle . 
~~\query
\end{equation}
The position vector is defined in light-front coordinates with
\begin{equation}
   y^\mu = (y^+,y^-,\T{y}) 
= \left( \frac{1}{\sqrt2}(y^0+y^3), \frac{1}{\sqrt2}(y^0-y^3), y^1, y^2 \right).
\end{equation}
These definitions of $P_i(x,\T{k})$ are of a so-called ``unintegrated
pdf''.  For the parton model for DIS we need the integral over all
$\T{k}$:
\begin{equation}
  \label{eq:pdf.def1b}
  f_i(x) =
        \int \frac{dy^-}{4\pi}
           \e^{-ixp^+y^- }
        \langle p| \, 
           \bar\psi_i(0,y^-, \T{0}) \gamma^+ \psi_i(0)
         \, |p\rangle . 
~~\query
\end{equation}
A many-body physicist (\eg, \cite{response}) would bring in the
concept of a ``response function''.  However, the definitions of pdf's
are tailored to their use in factorization theorems for
ultra-relativistic scattering.  Thus there are some interesting
differences compared with condensed matter or nuclear physics, whose
exploration deserves a separate discussion.

One problem with the above definitions is that they are not
gauge-invariant.  A second problem for the integrated pdf $f_i$ is
that it is UV divergent: the integral over $\T{k}$ diverges at large
$\T{k}$, as can be readily demonstrated from low-order Feynman graphs.
There is a third problem, which is more difficult to explain, but
which is at the root of the most interesting QCD issues.  This
concerns the divergences that arise when one solves the gauge
invariance in the most natural way, by applying the definitions in the
light-cone gauge $A^+=0$; these divergences are associated with the
$1/k^+$ singularities in the gluon propagator.  Essentially the same
divergences arise when the definition (\ref{eq:pdf.def1b}) is made
gauge-invariant in the natural way, by inserting Wilson lines in the
appropriate light-like direction.  If QCD were a superrenormalizable
theory with a scalar gluon, none of these problems would arise.

The primary topic of this article is to explain how these problems are
to be solved, so that a correct definition of pdf's can be made.  A
correct definition is one that allows a valid factorization theorem to
be derived correctly.  So problems with the definitions are correlated
with complications in the derivation of factorization.

\subsection{Renormalized operators: a valid definition of integrated pdf's}

First, let us consider the UV divergences in integrated pdf's.

It is important to remember that there are (at least) two kinds of
factorization theorem.  The first are the classical ones
\cite{ESW,CSS.review,CTEQ.handbook}, like Eq.\ (\ref{eq:DIS.factn});
they involve integrated pdf's.  The second kind are for processes such
as the Drell-Yan process at low transverse momentum \cite{CSS.qt};
these use the unintegrated pdf's.  Another example of a process of the
second kind is semi-inclusive DIS (SIDIS) when the transverse momentum
of a final-state hadron is measured: this process needs not only
unintegrated pdf's but also the corresponding unintegrated
fragmentation functions.  Unintegrated pdf's have found important uses
in treatments of polarized scattering \cite{SSA,SSA.JCC}.

The factorization theorems we are concerned with are those that have
been proved, or at least stated, for the whole leading power behavior
of cross sections.  This is in contrast to the many interesting
results that arise from a leading-logarithm analysis of perturbation
theory.

For fully inclusive DIS, the general structure of leading regions is
symbolized in Fig.\ \ref{fig:handbag}(b).  The consequences for
factorization are as follows.  First, there may be extra collinear
gluons exchanged between the proton-collinear subgraph and the
hard-scattering subgraph.  These gluons are allowed for
\cite{BL,fact.Wilson.lines,CSS1,CSS2} when integrated pdf's are
defined with a Wilson line between the quark and antiquark fields; the
resulting pdf's are gauge invariant.  A second complication is that
hard scattering subprocess can be arbitrarily more complicated than
the Born graph used in the handbag diagram.  However, apart from the
gluons that give the Wilson line factor, the hard scattering has the
minimum possible number of external parton lines, to avoid losing a
power of $Q$.  A single graph can have many different regions, and
this is correlated with the fact that the integrated pdf's as defined
below are UV divergent.  However, the end result is that a valid
factorization theorem holds if finite pdf's are defined by ordinary UV
renormalization of their UV divergences, and if the hard scattering
coefficients are constructed with suitable subtractions.  In that case
we have a valid pdf defined by \cite{fact.Wilson.lines,CS1}
\begin{equation}
  \label{eq:pdf.def2}
  f_i(x, \mu) =
        \int \frac{dy^-}{4\pi}
           \e^{-ixp^+y^- }
        \langle p| \, 
           \bar\psi_i(0,y^-, \T{0}) W[y,0] \gamma^+ \psi_i(0)
         \, |p\rangle_R . 
\end{equation}
The lack of the \query{} symbol indicates that this is a valid
definition, as far as I know.  Here $W[y,0]$ indicates a Wilson line
(\ie, a path-ordered exponential of the gluon field) along the
light-like straight line from the point $0$ to $(0,y^-,\T{0})$.  The
subscript `$R$' indicates that the operator is renormalized.  The
renormalization group equations for the dependence on the
renormalization scale $\mu$ of pdf's defined in this fashion are just
the well-known DGLAP equations.

However, as we will see in Sec.\ \ref{sec:UV.div}, the presence of UV
divergences removes the possibility of literally interpreting these
pdf's as number densities.  The factorization formulae merely permit
them to be used \emph{as if} they are number densities, since the
factorizations have the form of pdf's convoluted with a short-distance
cross section.

Another possible method of dealing with the UV divergences in the
integral over \emph{all} $\T{k}$ is to define an integrated pdf by an
integral over a restricted range $|\T{k}| < Q$, as advocated by
Brodsky and his collaborators \cite{BL,lc.div}:
\begin{equation}
  \label{eq:pdf.def3}
  f_i(x, Q) =
        \int_{|\T{k}| < Q} d^2\T{k} \, P_i(x,\T{k}).
~~\query
\end{equation}
As we will see in Sec.\ \ref{sec:UV.div}, the need for renormalizing
the quark fields still gives difficulties with a number
interpretation.  More importantly, this definition has divergences ---
see Sec.\ \ref{sec:lcg.div} --- associated with the use of the
light-cone gauge (or of the corresponding light-like Wilson lines),
contrary to the case when the integral is over all $\T{k}$, as in Eq.\ 
(\ref{eq:pdf.def2}).  Observe that the DGLAP scale dependence with
definition (\ref{eq:pdf.def3}) arises not only from the explicit upper
cutoff on $\T{k}$ but also from the anomalous dimension of the quark
fields in the definition of the unintegrated density $P_i(x,\T{k})$.

\subsection{Light-cone gauge}

Fig.\ \ref{fig:handbag}(b) has extra exchanged collinear gluons
compared with the handbag diagram.  One way of treating them \cite{BL}
is to use light-cone gauge, $A^+=0$, where these gluons are
power-suppressed.  Correspondingly, the Wilson line $W[y,0]$ in Eq.\ 
(\ref{eq:pdf.def1b}) is unity in this gauge, so that one can try to
define pdf's by applying the non-gauge-invariant definitions
(\ref{eq:pdf.def1}), (\ref{eq:pdf.def1a}), and (\ref{eq:pdf.def1b}) in
the $A^+=0$ gauge.

This is in fact the natural gauge for implementing light-front
quantization, so that these definitions give an elegant interpretation
of abstract operator definitions like Eq.\ (\ref{eq:pdf.def2}), as
expectation values of number operators.

Now, it is true that the divergences should cancel in a
gauge-invariant physical quantity, like a cross section.  So it could
be argued that the divergences in theorists' constructs like pdf's are
not really relevant.  However the cancellation of divergences only
holds if the physical quantity is computed exactly or in some
particular given order of perturbation theory.  Unfortunately,
factorization provides approximations that {\em mix} different orders.
For example, the result of a RG improved calculation of DIS structure
function has the schematic form
\begin{equation}
  F(Q) = f(\mu_0) 
         \exp\left[\int_{\mu_0}^Q \gamma(\alpha_s(\mu)) \, d\mu/\mu\right]
         C(Q/\mu) .
\end{equation}
The value of the pdf $f(\mu_0)$ is obtained from fits to data; it must
be considered as exact, complete with all non-perturbative
contributions.  Both the evolution kernel $\gamma$ and the hard scattering
coefficient $C$ are computed in perturbation theory truncated to some
low order.  The power of RG methods is that they show how make these
approximations correctly and systematically.

The truncations of perturbation theory in the hard scattering factor
and in the anomalous dimension in the exponent mean that an
approximation to the structure function\footnote{Please note that I
  maintain a strict and pedantic distinction between the concept of a
  structure function and a pdf.  A structure function is a property of
  a (measurable) cross section, while a pdf is a theorist's construct,
  a useful tool for the theoretical analysis and prediction of
  structure functions, etc.  } is not made at a single fixed order of
perturbation theory.  Therefore the use of factorization methods
depends critically on all the factors being individually finite.

\subsubsection{UV divergences}
\label{sec:UV.div}

To maintain a strict number interpretation of the pdf's, it is
essential not only that the pdf's defined by equations like Eq.\ 
(\ref{eq:pdf.def1a}) and (\ref{eq:pdf.def3}) are finite, but also that
the creation and annihilation operators have the standard commutation
relations:
\begin{equation}
  \label{eq:CR}
  \left[{a_i}^{\dag}(xp^+,\T{k},\lambda),
        {a_j}(x'p^+,\T{k'},\lambda')
  \right]
  =
  \delta_{ij} \delta_{\lambda\lambda'}
  (2\pi)^3 2x \delta(x-x') \delta^{(2)}(\T{k}-\T{k'}).
\end{equation}
These commutation relations follow, according to the principles of
light-front quantization, from the canonical commutation relations for
the bare field operators.\footnote{It would be useful to verify that
  these commutation relations do indeed follow from properties of the
  Heisenberg fields, as defined perturbatively by ordinary Feynman
  rules.  See Jaffe's work \cite{Jaffe} for some results in this area.
  }

The important point is that it is the {\em bare} fields, not the {\em
  renormalized} fields that obey canonical commutation relations.  On
the other hand, finite matrix elements of fields are those with
renormalized, not bare fields.  Thus the ${a_i}^{\dag}$ and ${a_i}$
operators in the definitions of the pdf's must be those obtained from
Fourier transforms of the renormalized fields.  Since the bare and
renormalized fields generally differ by an infinite factor, we must
conclude that the pdf's differ from actual number densities by
infinite factors. These considerations are entirely separate from the
issues recently discussed by Brodsky et al.\ under the title
``Structure functions are not parton probabilities''
\cite{parton.probs}.

As Brodsky et al.\ \cite{lc.div} explain, the UV problems can be
evaded by choosing to work with a large UV cutoff.  To avoid changing the
physics, this cutoff must be much larger than all experimental energy
scales.  But the continuum limit cannot be taken.

From the point of view of factorization theorems, the issue of UV
divergences is irrelevant.  All that matters is that one has some well
defined quantities that are labeled as pdf's, and in terms of which
useful factorization theorems are valid.

\subsubsection{Light-cone gauge divergences}
\label{sec:lcg.div}

Harder issues arise because of some well-known problems with the
light-cone gauge.  These cause divergences beyond those associated
with renormalization.  Essentially identical problems arise if the
pdf's are defined gauge invariantly with light-like Wilson lines.

General results that show that these definitions give divergences in
\emph{unintegrated} pdf's were obtained by Collins and Soper
\cite{CS1,CS2}.  They defined parton densities in an axial, or planar,
gauge $n\cdot A=0$ with a {\em non}-light-like gauge fixing vector
$n^\mu$.  Then they derived an equation for the gauge dependence, and
the solution of the equation gives a singular result in the
light-cone-gauge limit.  The result is valid at least to all orders of
perturbation theory.  Although the light-cone-gauge divergences cancel
in the integrated pdf's, they do not cancel in pdf's integrated up to
some finite limit.  Thus the cutoff method of defining an integrated
pdf, as in \cite{lc.div} Eq.\ (\ref{eq:pdf.def3}), cannot be applied
when the light-cone-gauge definition is used for the unintegrated pdf.

The technical derivation of the Collins-Soper equation was actually
only given for unintegrated fragmentation functions.  They and Sterman
\cite{CSS.qt} stated the corresponding result for pdf's, but did not
give an actual proof.  This equation has proved very useful in the
analysis of transverse momentum distributions\footnote{ A recent
  example is the analysis of transverse momentum distributions for the
  Drell-Yan process by Landry, et al.\ \cite{DY.qt}, where references
  to previous work can be found.  }.

To verify the existence of a divergence, we now examine a one-loop
calculation of the density of quarks in a quark\footnote{ At first
  sight, this concept appears paradoxical.  There are two quantitative
  definitions associated with the word ``quark''.  The first
  corresponds to a state that is created by a light-front creation
  operator, while the second corresponds to a one-particle energy
  eigenstate.  }.  The results can readily be obtained with the aid of
the Feynman rules in \cite{CS1}.  Perturbation-theory divergences
associated with the masslessness of the gluon in QCD are irrelevant to
our current purpose, as are the dimension of space-time and the
non-abelian nature of the gauge group.  So we will work with a gluon
of nonzero mass $m_g$, which is consistent if the gauge group is
abelian.  We will also have a nonzero quark mass $m$.  Thus we avoid
actual IR and collinear divergences. We will use a space-time
dimension $4-\epsilon$, so that the model is superrenormalizable; then no
divergent wave-function renormalization of the fields is needed.

The lowest-order value is just a delta function: $P_{0} = \delta(1-x)
\delta^{(2-\epsilon)}(\T{k})$.  The subscript ``$0$'' in ``$P_0$'' just
indicates its order in perturbation theory, $\alpha_s^0$. The fact that
the lowest order pdf is a delta function means that a numerical
interpretation is given by integrating with an arbitrary smooth test
function $t(x,\T{k})$.  Thus we write:
\begin{equation}
  \label{eq:pdf.0}
  P_0[t] \equiv
  \int dx\, d^{2-\epsilon}\T{k} \,
      t(x,\T{k}) \, P_0(x,\T{k})
  = t(0,\T{0}).  
\end{equation}
Treating the parton densities as generalized functions, to be
``integrated with a test function'' (to use the common abuse of
mathematical terminology), will enable us to perform a proper analysis
of the divergences in one-loop order and of their cancellation or lack
of it.

The one-loop graphs for real gluon-emission graphs give
\begin{eqnarray}
   \label{eq:1R}
   P_{1R}(x, \T{k}) 
&=&
   \frac{g^2}{ 2(2\pi)^{3-\epsilon}}
   \left\{
     \frac{ \displaystyle\frac{4}{\strut 1-x} -2 -2x -\epsilon(1-x)}
          { k_T^2 + m_g^2x + m^2(1-x)^2 }
   \right.
\nonumber\\
&&\hspace*{2cm}
  + \left.
     \frac{ x(1-x) 
            \left[
                 -4m^2 - 2m_g^2 + m_g^2\epsilon
            \right]
          }
          { \left[ k_T^2 + m_g^2x + m^2(1-x)^2 \right]^2 }
  \right\}.
\end{eqnarray}
The one-loop virtual graphs are proportional to a delta function.  An
explicit calculation shows that the coefficient is exactly the
integral of the real emission graphs over all $x$ and $\T{k}$, but
with the sign changed:
\begin{equation}
  \label{eq:1V}
   P_{1V}(x, \T{k}) 
   = - \delta(1-x) \, \delta^{(2-\epsilon)}(\T{k})
     \int_0^1 d\alpha  \int d^{2-\epsilon}\T{l} \, 
       P_{1R}(\alpha, \T{l}).
\end{equation}
Observe that the $4/(1-x)$ term in the first line of Eq.\ 
(\ref{eq:1R}) implies that $P_{1V}$ is divergent.  If the calculation
is done in light-cone gauge, the divergence is a direct consequence of
the $1/k^+$ singularity in the gluon propagator.  If the calculation
is done in Feynman gauge, the divergence arises from the corresponding
singularity in the Feynman rule for the Wilson line.

The divergence is caused by an endpoint singularity, so it cannot be
removed by changing the analytic prescription for the singularity of
the gluon propagator in light-cone gauge.  See Brodsky et al.\ 
\cite{lc.div} for another calculation of the same divergence.

Given that divergences often cancel between real and virtual graphs,
we should add the real and virtual contributions, which we must do in
the sense of generalized functions, \ie, with an integral with a test
function:
\begin{eqnarray}
   \label{eq:1}
  &&\hspace*{-1cm}
  \int dx\, d^{2-\epsilon}\T{k} \,
      t(x,\T{k}) \, P_1(x,\T{k})
~=
\nonumber\\
&& \hspace*{-1cm}
=~
   \frac{g^2}{ 2(2\pi)^{3-\epsilon}}
  \int_0^1 dx  \int d^{2-\epsilon}\T{k}
    \left[ t(x,\T{k}) - t(1,\T{0}) \right] 
\times
\nonumber\\
&&\hspace*{-0.5cm}
\times
   \left\{
     \frac{ \displaystyle\frac{4}{\strut 1-x} -2 -2x -\epsilon(1-x)}
          { k_T^2 + m_g^2x + m^2(1-x)^2 }
   +
     \frac{ x(1-x) 
            \left[
                 -4m^2 - 2m_g^2 + m_g^2\epsilon
            \right]
          }
          { \left[ k_T^2 + m_g^2x + m^2(1-x)^2 \right]^2 }
  \right\}.
\end{eqnarray}
Since we are manipulating divergent integrals, we should actually
apply a regulator (for example, we could apply a cut off on the plus
momenta in the theory) in order to derive this formula correctly.

If the test function is replaced by a function independent of $\T{k}$,
then the divergence does cancel, because the factor $t(x,\T{k}) -
t(1,\T{0})$ becomes $t(x) - t(1)$, which is zero at $x=1$.  This
result was given by Collins and Soper \cite{CS1}. Thus the integrated
parton density does exist, in the sense of generalized functions.

However if the test function is $\T{k}$ dependent, then the divergence
does not generally cancel.  In particular, if the integrated pdf
were defined with an cutoff on transverse momentum, as in Eq.\
(\ref{eq:pdf.def3}), it would have an uncanceled divergence.

\subsubsection{Interpretation of light-cone gauge divergences}

Examination of the derivation of the above results shows that the plus
component of the gluon's momentum is $(1-x) p^+$ [or $(1-\alpha)
p^+$].  So at first sight the divergence at $x \to 1$ is a soft
divergence, similar to the ones in QED.  But since we have nonzero
gluon and quark masses, this is not a correct interpretation; the
divergence exists for any value of $\T{k}$.

In fact the divergence comes from a region where the gluon rapidity
goes to minus infinity.  That is, it is from where the gluon is going
infinitely fast in the direction of the outgoing quark jet (if we
consider the corresponding DIS kinematics in the Breit frame, as in
Fig.\ \ref{fig:DIS}).  It therefore corresponds to a region of
momentum that has nothing to do with the region of momenta collinear
to the proton that was associated with the pdf in deriving
factorization.  We must therefore say that we have an inappropriate
definition of a pdf.

\section{Correct definitions (I hope) of unintegrated pdf's}
\label{sec:correct}

The derivation of factorization involved making an approximation
appropriate for certain regions of momenta, and the definition of the
pdf arose from extrapolating this definition beyond the region of
validity of the approximation.  

\subsection{Options}

Since there are divergences in the pdf, as we have seen, the
definition must be modified to incorporate some kind of cutoff on
gluon rapidity or some kind of generalized renormalization.  The
cutoff must apply to gluons in virtual graphs.  However, we should not
implement the cutoff in the Lagrangian of QCD.  For then the cutoff
would depend on which hadron in a process we are considering, whereas
the QCD Lagrangian is supposed to describe all possible processes on
all possible momentum scales.  Moreover we would simultaneously need
opposite cutoffs for different parts of the same process.  For
example, we would need a cutoff on gluons of {\em negative} rapidity
to define pdf's in the target hadron in DIS, but we would a cutoff on
gluons of {\em positive} rapidity to define the fragmentation
function.

So the cutoff belongs in the definition of the operator whose
expectation value is the parton density.  There are several
possibilities, including:
\begin{itemize}
\item Use the non-gauge-invariant definition Eq.\ 
  (\ref{eq:pdf.def1a}), but choose the gauge to be an axial/planar
  gauge $n\cdot A=0$ with a {\em non}-light-like gauge fixing vector
  $n^\mu$, as proposed by Collins and Soper \cite{CS1}.  This is
  basically correct, but it does not contain the analytic properties
  associated with the directions of the Wilson lines that a correct
  Feynman-gauge derivation of factorization would naturally give.  See
  \cite{SSA.BHS,SSA.JCC} for recent work on spin-dependent processes
  where the process-dependent direction of the Wilson line is critical
  to getting the correct relative signs for single spin asymmetries,
  which are non-universal between different processes.
  
\item This problem can be overcome by inserting Wilson lines in
  non-light-like directions \cite{Sud}.  It is important that the
  derivation of a valid factorization produces certain constraints of
  the directions of the Wilson lines.  It is not sufficient, for
  example, simply to choose the Wilson line to be along the straight
  lines joining the quark and antiquark field.
  
\item Use the definition with light-like Wilson lines, but multiply by
  a suitable gauge-invariant factor that cancels the divergence.  This
  was suggested by Collins and Hautmann \cite{CH} on the basis of a
  one-loop calculation.  This gives a kind of generalized
  renormalization with the renormalization factors being certain
  vacuum expectation values of Wilson line operators, with a mixture
  of light-like and non-light-like lines.  Their definition appears to
  follow almost uniquely from the structure of the asymptotics of
  one-loop graphs.

\end{itemize}
I propose that either of the last two methods is appropriate and
valid, with the third method being the more elegant mathematically.
Observe that simply using light-like Wilson lines without a further
generalized renormalization factor does not remove the divergences.

All of these definitions involve two auxiliary parameters, a
renormalization scale $\mu$ and an effective cutoff on gluon
rapidity.  Collins and Soper \cite{CS1} introduced a parameter
$\zeta$, which can be interpreted as $p^2\cosh^2\Delta y$, where
$\Delta y$ is the difference of rapidity between the target hadron and
the gluon rapidity cutoff.

These extra parameters mean that the pdf's suitable for applying
factorization depend on the energy of the process, thereby endangering
universality of pdf's.  Universality, which allows the pdf's to be
used phenomenologically, is regained with the aid of evolution
equations for the dependence on the auxiliary parameters. These were
obtained by Collins, Soper, and Sterman \cite{CSS.qt,CS1,CS2}. Their
equations are very different to the normal DGLAP equations, even
though some of the physics content is related. Developments of these
equations for understanding the large $x$ behavior of pdf's have been
obtained by Sterman \cite{large.x}.

\subsection{Non-light-like Wilson line}

The first option is to define the unintegrated pdf's gauge-invariantly
with non-light-like Wilson lines.  Just as in Eq.\ 
(\ref{eq:pdf.def1b}), the pdf is a Fourier transform:
\begin{equation}
  \label{eq:pdf.ft}
  P_i(x,\T{k}, \zeta, \mu) =
        \int \frac{dy^-\, d^2\T{y}}{16\pi^3}
           \e^{-ixp^+y^- + i \T{k}\cdot\T{y} } \,
        \tilde{P}_i(y^-,\T{y}, \zeta, \mu) ,
\end{equation}
but where the quark correlation function has Wilson line factors:
\begin{equation}
  \label{eq:pdf.def4}
  \tilde{P}_i(y^-,\T{y}, \zeta, \mu) =
        \langle p| \, 
           \bar\psi_i(0,y^-, \T{y}) \, W_y(u)^{\dag} \,
           I_{u;y,0} \,
           \gamma^+ \, 
           W_0(u) \, \psi_i(0) \,
         |p\rangle_R . 
\end{equation}
Here $W_y(u)$ denotes the following Wilson line operator
\begin{equation}
  \label{eq:WL.def}
  W_y(u) = P \exp\left[
             -ig_{(0)} \int_0^\infty d\lambda \,
                u^\mu \, A_\mu^{(0)}(y+\lambda u)
           \right] ,
\end{equation}
where the line is from the position $y^\mu$ going to infinity in the
direction $u^\mu$.  To make the definition exactly gauge-invariant, a
Wilson line $I_{u;y,0}$ at infinity is needed \cite{BJY} to join $W_y(u)$
and $W_0(u)$.  In Feynman gauge this line at infinity  can be replaced
by unity. 
Note the following
\begin{itemize}

\item This gives a gauge invariant definition of the unintegrated
  pdf, with the coupling and the gauge field being the bare
  quantities, as indicated by the sub/superscript ``$(0)$''.
  
\item The vector $u^\mu$ that sets the direction of the Wilson line
  must not be light-like.  The parton density therefore depends on the
  variable $\zeta = (p \cdot u)^2/u^2$.
  
\item According to \cite{CSS.qt,CS2}, the proof of a factorization
  theorem requires that $u^\mu$ should be approximately ``at rest'' in
  the center of mass of the hard scattering.  Thus the Wilson lines
  are used, roughly speaking, to separate the gluons associated with
  different jets.
  
\item Hence $\zeta \sim s$, the square of the center-of-mass energy.
  Refs.\ \cite{CSS.qt,CS2} give an equation for the $\zeta$
  dependence of the pdf.
  
\item Strict gauge invariance requires that the two Wilson lines be
  connected by a link at infinity, $I_{u;y,0}$, as Belitsky, Ji, and
  Yuan \cite{BJY} have observed.  When Feynman gauge is used, at least
  in simple calculations, this extra link does not contribute.  But in
  light-cone gauge, the link at infinity is the only part of the
  Wilson line that contributes, and is critical in obtaining correct
  values for time-reversal-odd pdf's.
  
\item Whether $u^\mu$ points to the future or past depends on the
  process.  For DIS-type process a future pointing line is needed,
  which corresponds to final-state interactions in the Breit frame.
  For DY-type processes a past-pointing line, associated with
  initial-state interactions in the center-of-mass, is needed.  Time
  reversal invariance can be used to relate the two cases
  \cite{SSA.JCC}, and gives a reversal of sign for time-reversal-odd
  pdf's.

\item There are UV divergences associated with the quark-Wilson-line
  vertex that must be renormalized away by conventional methods.  This
  is indicated by the subscript ``$R$'' in Eq.\ (\ref{eq:pdf.def4}).

\end{itemize}

\subsection{Light-like Wilson line with generalized renormalization}

The definition Eq.\ (\ref{eq:pdf.def4}) is a gauge-invariant
transcription and correction of the planar-gauge definition by
Collins and Soper \cite{CS1}.  However, the non-light-like Wilson
lines complicate explicit Feynman graph calculations, compared with
those that use light-like Wilson lines.  There is also a mathematical
problem.  The exact evolution equation has an inhomogeneous term that
is power-suppressed and that is ignored in applications.  It would be
preferable to have an exactly homogeneous equation.

\begin{figure}
\centering
    \includegraphics[scale=0.25]{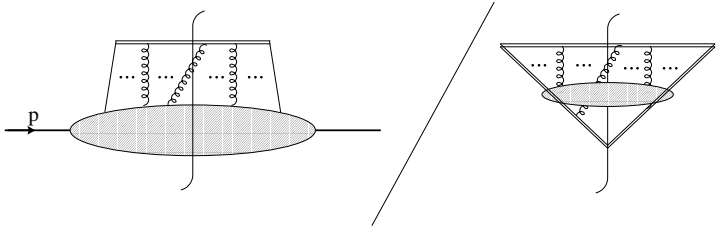}
    \caption{Diagrammatic interpretation of Eq.\ (\ref{eq:pdf.def5}).
      This is ordinary pdf in coordinate space (with a light-like
      Wilson line) divided by the vacuum expectation value of a
      certain pure Wilson line operator.  Any number of gluon lines
      can join the lower blob to the top Wilson line in the numerator.
      Any number of gluon lines can join the Wilson lines in the
      denominator. }
    \label{fig:def5}
\end{figure}
I therefore propose an alternative definition.  A light-like Wilson
line is used in Eq.\ (\ref{eq:pdf.def4}) and the consequent
divergences are canceled by a kind of generalized renormalization
factor.  From the work of Collins and Hautmann \cite{CH}, I conjecture
that a valid definition is
\begin{equation}
  \label{eq:pdf.def5}
  \tilde{P}^{\rm alt}_i(y^-,\T{y}, \zeta, \mu) =
  \frac{
        \langle p| \, 
           \bar\psi_i(0,y^-, \T{y}) \, W_y(n)^{\dag} \,
           I_{n;y,0} \, \gamma^+ \,
           W_0(n) \, \psi_i(0) \,
        |p\rangle_R 
  }{
        \langle 0| \, 
           W_y(n)^{\dag} \, W_y(u') \,
           I_{n;y,0} \, I_{u';y,0}^{\dag} \,
           W_0(n) \, W_0(u')^{\dag} \,
        |0\rangle_R     
  }. 
\end{equation}
This is illustrated in Fig.\ \ref{fig:def5}.  The Wilson line
associated with the quark fields is now in an exactly light-like
direction $n^\mu=(n^+,n^-,\T{n})=(0,\pm1,\T{0})$, which is future
pointing or past pointing depending on the process \cite{SSA.JCC}.
The vector ${u'}^\mu$ in the Wilson line in the denominator plays the
role of the vector $u^\mu$ in the previous
definition.\footnote{However there will probably be different signs in
  its plus and minus components \cite{CH,Metz}, to allow an exactly
  correct derivation of factorization.  This point is rather subtle,
  however. }
The pdf is defined by taking $n^\mu$ initially non-light-like,
computing the ratio in Eq.\ (\ref{eq:pdf.def5}), and then taking the
limit that $n^\mu$ is light-like.  One-loop calculations
\cite{CH,Metz} indicate that this can be implemented by a subtraction
method.

\section{Summary}

\subsection{Integrated pdf's}

\begin{itemize}

\item Integrated pdf's can be defined by formulae like Eq.\ 
  (\ref{eq:pdf.def2}).  They involve an integral over all transverse
  momentum, strictly out to infinity.  The resulting UV divergences
  are removed by conventional renormalization.  
  
\item Factorization theorems containing these pdf's have been proved
  for various inclusive processes \cite{CSS1,CSS2,Bodwin}.

\item In this formalism, the DGLAP equations are exactly the
  renormal\-iza\-tion-group equations for the pdf's

\item Divergences associated with the use of light-cone gauge or of
  light-like Wilson lines cancel after the integral to infinite
  transverse momentum
  
\item The presence of UV divergences prevents a literal number
  interpretation of these pdf's. Even so, momentum and quark-number
  sum rules are exact, provided a suitable renormalization scheme is
  used, \eg, \MSbar.  The proof \cite{CS1} relates the relevant
  moments of the pdf's to matrix elements of conserved Noether
  currents.

\end{itemize}

\subsection{Unintegrated pdf's}

\begin{itemize}

\item The most obvious definitions of unintegrated pdf's, \ie, of
  transverse-momentum-dependent pdf's, are plagued with divergences
  associated with the use of light-cone gauge, or with the
  corresponding light-like Wilson lines.

\item Some kind of cut-off or generalized renormalization must be
  apply to remove these divergences.

\item Two possible definitions are proposed, in Eqs.\
  (\ref{eq:pdf.ft}), (\ref{eq:pdf.def4}), and (\ref{eq:pdf.def5}).
  
\item One of these is a gauge invariant transcription of the
  Collins-Soper definition \cite{CS1}; the other is a potential
  improvement better adapted to subtractive methods of proof and
  calculation.
  
\item Given the auxiliary steps needed to define the unintegrated
  pdf's, it does not appear that they have a literal number
  interpretation. One uses them in factorization theorems as if they
  are number densities, but the strict number interpretation is not
  needed.
  
\item Given the extra cutoff needed on gluon rapidity, one should
  \emph{not} expect that the integral over transverse momentum of an
  unintegrated pdf should be exactly the corresponding integrated pdf.
  The relation between the two involves a non-trivial but
  perturbatively calculable coefficient \cite{CS1,CS2}. 
  
\item The existence of the extra cutoff implies that phenomenological
  applications should use an evolution equation \cite{CS1,CS2,CSS.qt}
  for the dependence of the unintegrated pdf on this cutoff.

\end{itemize}

\subsection{What needs to be done}

The reader who tries to find detailed justification for many of the
statements in this article will probably be quite frustrated.  If
nothing else, the literature on the subject is very fragmented.  Many
results which are reasonably clear to experts in the field, \eg, as
minor modifications to previously existing results, are quite
unobvious to outsiders and newcomers.

Moreover, published definitions of pdf's taken literally often give
divergences.  Given the great importance of pdf's and factorization
theorems to the whole of high-energy physics, it is important to fully
systematize the subject.

In this paper I have attempted to make a contribution to this
systematization by presenting what I believe to be complete
definitions of pdf's for quarks, with all known sources of divergences
explained and explicitly overcome by suitable prescriptions for
(generalized) renormalization and/or cutoff.  These should, of course,
be useful for the precise formulation of higher-order perturbative QCD
calculations.  But they should also be for providing a fully and
rigorously defined interface to models of non-perturbative hadronic
structure that are becoming increasingly important.

It is with pleasure that I dedicate this work to Jan Kwieci{\'n}ski.

\section*{Acknowledgments}

I would like to thank S. Brodsky, Y. Dokshitzer, A. Metz, V.
Pandharipande, M. Paris, T. Rogers, and X. Zu for useful discussions.
This work was supported in part by the U.S. Department of Energy under
grant number DE-FG02-90ER-40577.


\end{document}